\documentclass[twocolumn,secnumarabic,amssymb, nobibnotes, aps, prd]{revtex4-1}
\usepackage{graphicx}
\usepackage{dcolumn}
\usepackage{bm}
\usepackage{comment}
\usepackage{epstopdf}
\usepackage{natbib}

\newcommand{\be}{\begin{equation}}
\newcommand{\ee}{\end{equation}}
\newcommand{\bea}{\begin{eqnarray}}
\newcommand{\eea}{\end{eqnarray}}

\begin{document}

\preprint{AIP/123-QED}

\title{Spin-orbit coupling and transport of strongly correlated two-dimensional systems}
\thanks{email:jianhuang@wayne.edu}

\author{Jian Huang}
\affiliation{%
Department of Physics and Astronomy, Wayne State University, Detroit, MI 48201, USA\\}%
\author{L. N. Pfeiffer}%
\author{K. W. West}%
\affiliation{%
Department of Electrical Engineering, Princeton University, Princeton, NJ 08544}%

\date{\today}

\begin{abstract}
{Measuring the magnetoresistance (MR) of ultraclean {\it GaAs} two-dimensional holes in a large $r_s$ range of 20-50, two striking behaviors in relation to the spin-orbit coupling (SOC) emerge in response to strong electron-electron interaction. First, in exact correspondence to the zero-field metal-to-insulator transition (MIT), the sign of the MR switches from being positive in the metallic regime to being negative in the insulating regime when the carrier density crosses the critical density $p_c$ of MIT ($r_s\sim 39$). Second, as the SOC-driven correction $\Delta\rho$ to the MR decreases with reducing carrier density (or the in-plane wave vector), it exhibits an upturn in the close proximity just above $p_c$ where $r_s$ is beyond 30, indicating a substantially enhanced SOC effect. This peculiar behavior echoes with a trend of delocalization long suspected for the SOC-interaction interplay. Meanwhile, for $p<p_c$ or $r_s>40$, in contrast to the common belief that a magnet field enhances Wigner crystallization, the negative MR is likely linked to enhanced interaction.}
\end{abstract}

\pacs{Valid PACS appear here}
\maketitle


Electron systems governed by strong Coulomb interaction manifest unique charge transport behaviors characterized by the collective modes such as those in the Wigner crystal (WC)~\cite{wc}, fractional quantum Hall states (FQHS)~\cite{tsui82}, and the zero-field metal-to-insulator transition (MIT)~\cite{mit}. There is an important interaction-driven effect associated with the spin-orbit coupling (SOC) which has attracted a lot of interests due to its promising spintronic applications. SOC studies often utilize systems that lack inversion symmetries, both bulk and structural inversion asymmetries (BIA and SIA), which give rise to Dresselhaus~\cite{dresselhaus} and  Rashba SOCs~\cite{rashba}. As a result, the spin degeneracy of the energy bands is lifted even in a zero magnetic field ($B$). The band splitting in response to controllable quantities, especially electron-electron interaction, is a fundamental effect that is not well understood. SOC-driven effects in weakly interacting systems are usually perturbations, i.e. those tested $p$-type two-dimensional (2D) {\it InGaAs, GaAs/AlGaAs} heterostructures with carrier densities $\sim2-5\times10^{11}$ cm$^{-2}$, where $r_s=E_{ee}/E_F=a/a_B\sim1-5$ and $k_F l\gg1$. $E_{ee}\sim e^2/\epsilon r$-Coulomb energy, $E_F=n\pi\hbar^2/m^*$-Fermi energy, $a$-average charge spacing, $a_B$-Bohr radius, $k_F$-Fermi wave vector, and $l$-mean free path. However, in a strongly correlated system, the SOC-interaction interplay modifies the exchange interaction and leads to more prominent effects~\cite{Altshuler,pudalovSOC,Giuliani}. For example, a diverging density of states (DOS) known as the Van Hove sigularity is expected in the limit of small wavevectors $k_{\parallel}$ (or very low carrier density). However, such effects have not been studied previously due to lack of access to strongly interaction-driven systems. Large $r_s=E_{ee}/E_F=a/a_b\gg 1$ in the absence of a magnetic field requires very low carrier concentrations for which the usual level of disorder renders an Anderson localization (or percolation transition). This study utilizes ultra-high purity two-dimensional (2D) holes in {\it GaAs} heterostructures and demonstrates enhanced SOC-driven effects in a truly interaction-driven limit.

In {\it p}-type (accumulation-type) GaAs heterostructures, SOC results in splittings of both the light hole (LH) and heavy hole (HH) bands. While the BIA-induced effect remains constant, the SIA (Rashba) contribution can be tuned cexternally, i.e. via a metal gate. For high carrier densities, BIA-induced splitting of the HH band is dominated by a $k_{\parallel}^3$ contribution ($k_{\parallel}$ is the in-plane carrier momentum). However, the situation becomes complicated when the carriers are sufficiently dilute. Apart from weakening the external electric field, reducing the gate bias modifies the confinement potential which affects the HH-LH separation and thus influences the Rashba coefficient ($\alpha$). Meanwhile, the effect of the LH-HH mixing (noparabolic dispersion) is expected to rise. Recently MR measurements on the $p$-dependence shows a moderately increasing splitting with reducing $k_{\parallel}$~\cite{Winkler2}, suggesting an enhancement of $\alpha$ possibly related to the LH-HH mixing. However, due to limited density range, such effects are not explored into the strongly correlated metal-to-insulator (MIT) transition regime~\cite{pudalovSOC} where larger enhancement is anticipated for small enough $k_{\parallel}$~\cite{Utah}.

We adopt undoped ultra-dilute two-dimensional (2D) holes in GaAs HIGFETs (hetero-junction-insulated-gate field-effect-transistors) in which the carrier density $p$ can be continuously tuned from 0.2 to $2\times10^{10}$ cm$^{-2}$ (or $r_s$ from 60 to 25). This allows a first probe to the MR across MIT with the critical density $p_c$ being only $4\times10^{9}$ cm$^{-2}$ (or $r_s\sim 39$ if an effective mass $m*=0.3m_0$ is assumed). The zero-field temperature ($T$) dependence of the conductance shows non-activated transport, excluding the domination of an Anderson-localization, which is a crucial indicator for an interaction-driven nature. For a decreasing $p$ from well above $p_c$ down to the proximity of $p_c$, the measured $p$-dependence of the MR captures a substantial resistivity variation (or correction) $\Delta\rho$ that corresponds to a nearly three-fold rise around $6\times10^{9}$ cm$^{-2}$. Moreover, crossing into an insulator at $p_c$, the positive sign of MR switches rapidly to negative, in contrast to the usual expectation of a WC stabilized by a $B$ field.  
\begin{figure}[t]
 \centering
 \includegraphics[totalheight=1.6in]{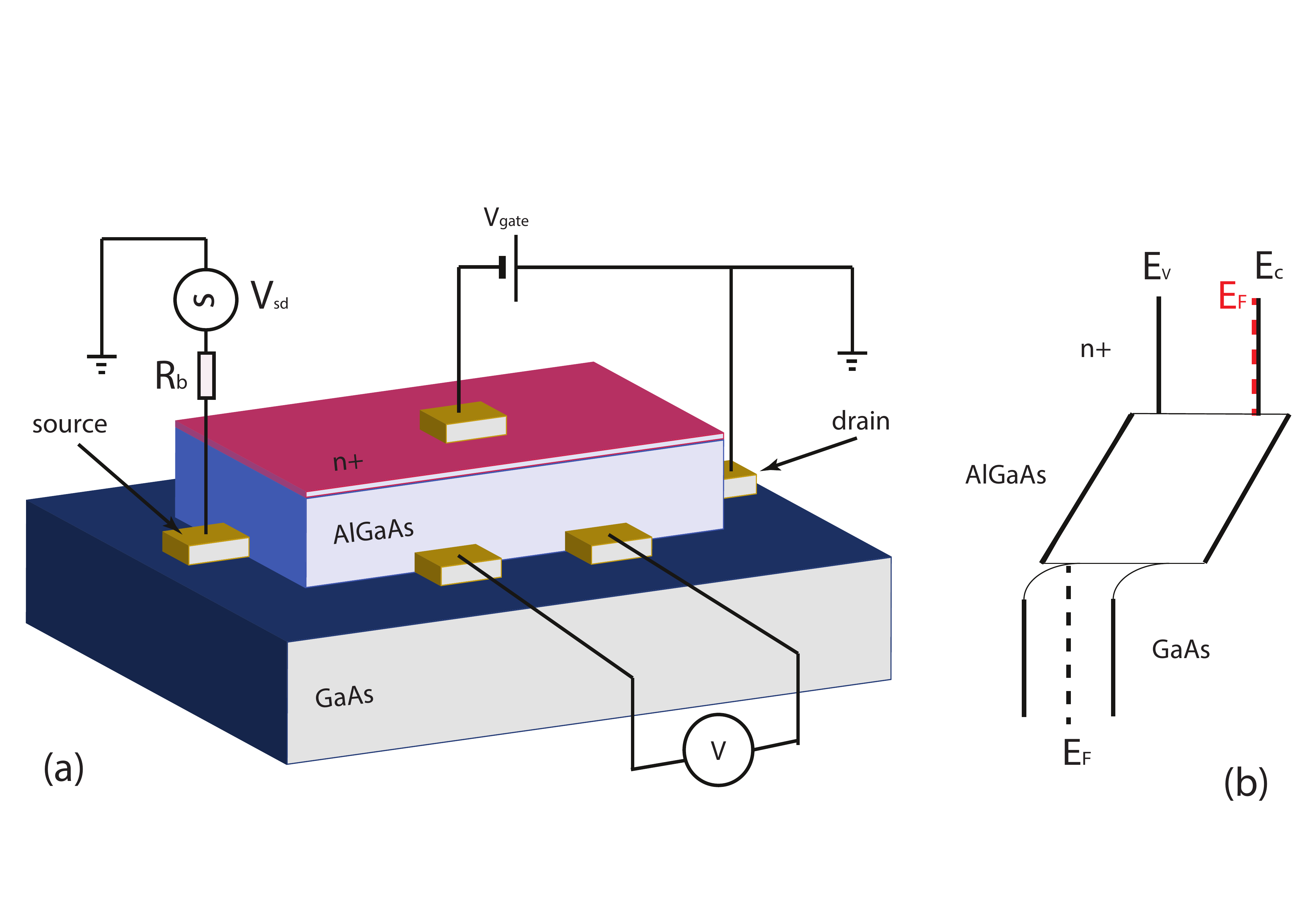}
 \caption{\label{fig:device} (a) Schematics for the HIGFET, contacts, and measurement setup. (b) Band structure under gate voltage bias.}
\end{figure}

The devices adopted are 2D holes in undoped $p$-channel $\langle100\rangle$ HIGFET~\cite{kane,noh,jian06} patterned into 1 mm $\times$ 6 mm Hallbar shapes. As illustrated in Fig.~\ref{fig:device} (a), the 2D holes are confined in the $\sim$10-30 nm triangular well at GaAs/Al$_{0.3}$GaAs hetero-interface. The Ohmic contacts are made with AuBe alloy annealed at 420$^o$C and the contacts to the top gate layer of $n^+$-GaAs are made with Cr/Au alloy without annealing. The top gate, a 30 nm heavily doped GaAs layer, and the 2D hole layer form a capacitor with 600 nm of dielectric AlGaAs layer in between. The 2D holes are, as shown in Fig.~\ref{fig:device} (b), only capacitively induced by gating the $n^+$ layer beyond the threshold voltage $V_c\approx$\,-1.3 V with respect to 2D channel. $p$ is tuned from $7\times10^{8}$ $cm^{-2}$ to $1.5\times10^{10}$ cm$^{-2}$~\cite{jian06} via varying $V_{gate}$ and is determined through quantum Hall measurements with a field sweep rate of 100 Gauss/min in a dilution refrigerator. 

We first present the $T$-dependence of the conductivity ($\sigma$) and resistivity ($\rho$) results for a range of $p$ from 2 to $18\times10^{9}$ $cm^{-2}$ in a zero-field. It is well known that hopping conductance is often assumed for all dilute systems. This means that, with small $E_{ee}\sim 1m$eV or 10 K and $E_F\sim300 m$K, disordered systems usually give in to unscreened (or poorly screened) disorders and undergoes Anderson Localization~\cite{anderson58}. As a result, one observes exponentially suppressed conductance by cooling: $\sigma\sim e^{-(T^*/T)^\nu}$. $\nu=1$ is for Arrhenius case at higher $T$ and $\nu=1/3$ and $1/2$ for variable-range hopping (VRH)~\cite{vrh1,vrh2}. Even with relatively clean systems, the Anderson scenario can still hold if the single particle localization length $\xi$ is exceeded by the average charge spacing $a=1/\sqrt{\pi p}$ at low enough $p$. Thus, ultrahigh purity systems are required so that interaction-driven nature persists even at the onset of a WC~\cite{wc1}, $r_s\sim 40$, which corresponds to $p\sim4\times10^{9}$ cm$^{-2}$ (or $1\times10^{9}$ cm$^{-2}$ for electrons). 

\begin{figure}[b]
 \centering
 \includegraphics[totalheight=2.1in]{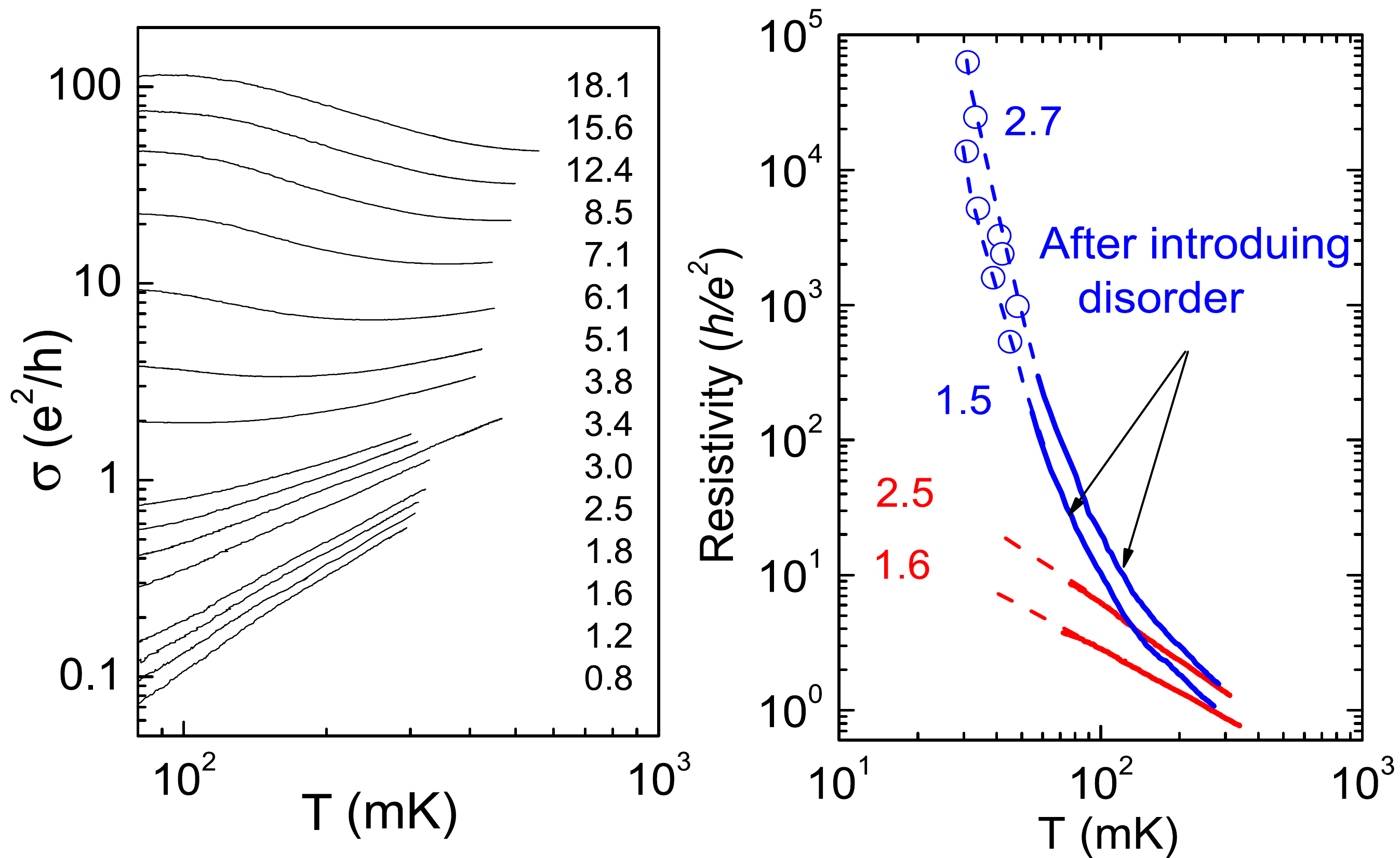}
 \caption{\label{fig:RT} (a) Conductivity $\sigma$ vs. $T$ in $log$ scales for various $p$ from 0.8 to 18$\times10^{9}$ $cm^{-2}$. (b) Comparison of the resistivity $\rho$ vs. $T$ (in $log$ scales)  for selected $p$'s before and after disorder is introduced. Dotted lines are guide for the eye.}
\end{figure}

As shown in Fig.~\ref{fig:RT}(a) in log-log scales, the apparent MIT occurs around $p_c\sim4\times10^{9}$ cm$^{-2}$ below which a striking non-activated power law behavior~\cite{jian06,jian12}, $\sigma\sim (T/T^*)^\gamma$, is observed in multiple tested samples. $p_c$ corresponds to $r_s\sim39$~\cite{wc1}, the anticipated onset point of a WC. Notice that the critical density obtained previously in more disordered systems varies substantially from sample to sample, i.e. $8\times10^{10}$ to $3\times10^{9}$ cm$^{-2}$ (or $3<r_s<15$) for electron devices and $8\times10^{10}$ to $8\times10^{9}$ cm$^{-2}$ (or $5<r_s<25$ for holes. However, the $p_c$ found in HIGFETs is significantly lower and varies little among many different samples: 3.8 to $4.4\times10^{9}$ cm$^{-2}$. The nonactivated power-law $T$-dependence on the insulating side contrasts the disorder-dominated hopping scenario. The exponent $\alpha$ varies from 1 to 2 with decreasing $p$. This power law $T$-dependence, which has been reported for both GaAs~\cite{jian12} and SiGe~\cite{Lai} systems, likely belongs to a strongly correlated liquid (i.e. a melted WC or glass) since the experimental temperature is above the WC melting temperature $T_{m}$. 

An important fact about the power-law $T$-dependence is its vulnerability to even a slight increase of disorder. Long-range disorders are introduced through LED illumination to the sample within the  same cooling cycle. The photons energy is approximately the band gap in {\it GaAs} and {\it AlGaAs}. There is at least 24 hours of wait time after the illumination before measurement resumes and the uniformity of the carriers are verified via the Hall measurement. The amount of disorder introduced is roughly controlled by the time duration of the light exposure, which is usually 0.5-2s, at a constant 0.05 $\mu$A current excitation. This causes $p_c$ to rise~\cite{jianlight}: i.e. $p_c$ is almost doubled after 0.5s of illumination and tripled after 1.5 s illumination. Meanwhile, as shown in the $log-log$ plot in Fig.~\ref{fig:RT}(b) for $p=1.5-2.8\times10^{9}$ cm$^{-2}$, $\rho(T)$ undergoes a qualitative change from the power-law (prior to the illumination) into an exponential law $\rho\sim e^{-(T^*/T)}$ of hopping, with a characteristic energy $T^*\sim200-400$ mK. $\rho$ jumps by orders of magnitude. This qualitative change of behaviors indicates that the power-law represents a different state likely due to strong correlation. We performed further work to confirm this by studying the exponent $\gamma$ of the power-law and found it scales with a dimensionless parameter $a/d$ with $a$ being the charge spacing, and $d$  the distance to the gate. This is due to the onset of a dipolar screening~\cite{jian-screen} and thus confirms the interaction-driven nature which is foundational to the following MR results.  

\begin{figure}[t]
 \centering
 \includegraphics[totalheight=2.4in]{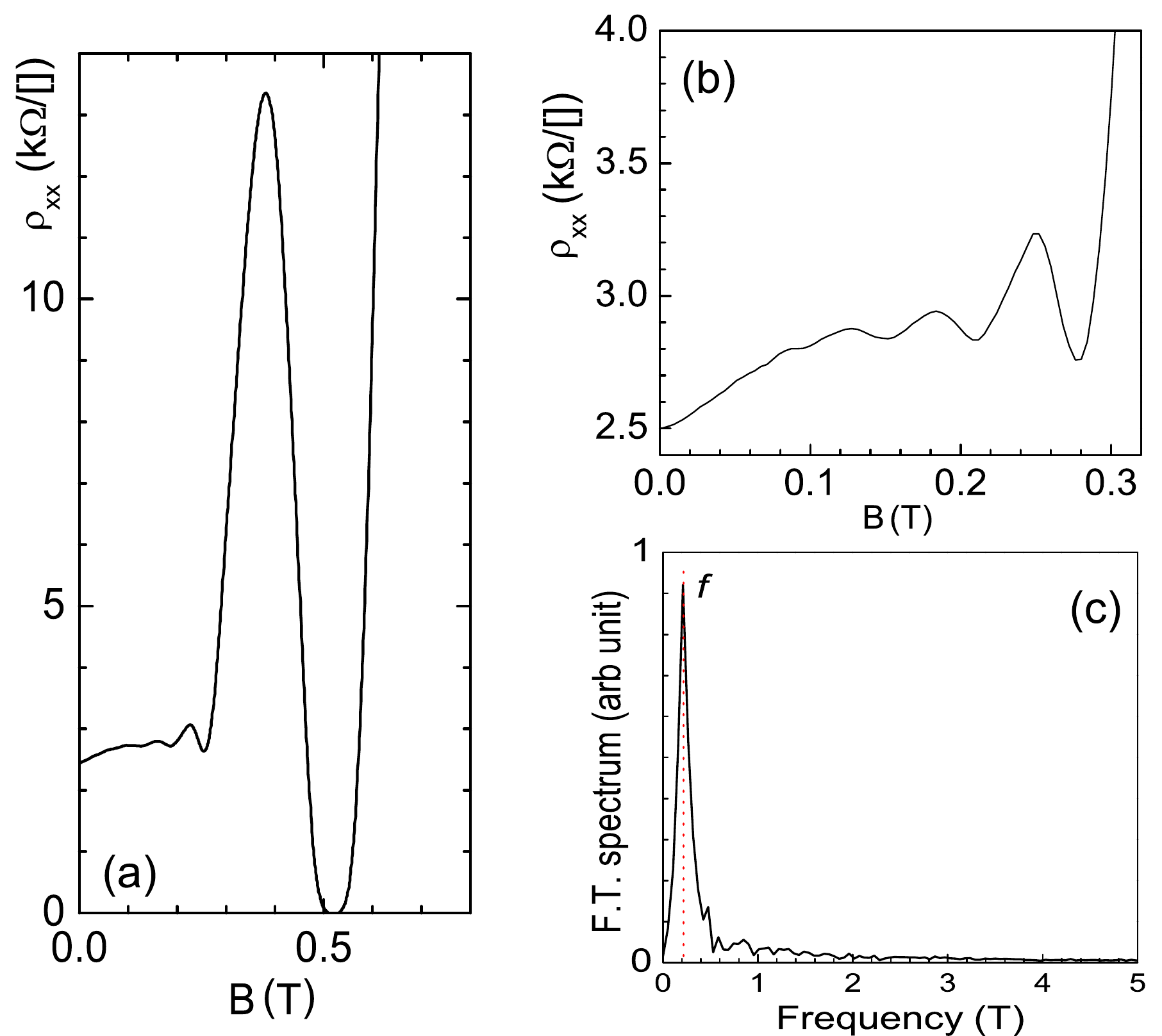}
 \caption{\label{fig:Rxx} (a) $\rho_{xx}$ vs. B for $p=1.2\times10^{10}$ cm$^{-2}$ at $T$= 45 mK. (b) SdH oscillations seen in a zoom-in view of (a) at low fields. Inset: Fourier spectrum of the shown $\rho_{xx}(B)$.}
 \vspace{-10pt}
\end{figure}

The longitudinal MR, $\rho_{xx}$ (B), shown in Fig.~\ref{fig:Rxx}(a), is for $p\sim1.2\times10^{10}$ cm$^{-2}$ (or $r_s\sim30$) with a carrier mobility of $\mu\sim300,000$ cm$^{2}/(V\cdot s)$ and $\rho_{xx}(0)=2.35$ k$\Omega/\sim0.1h/e^2$. $h/e^2$ is the quantum resistance. Shubnikov-de Haas oscillations (SdH) [Fig.~\ref{fig:Rxx}(b)] are observed between 0.05 to 0.25T before a substantial (re-entrant insulating) peak develops at $B=0.37$T proceeding the filling factor 1 (at $\sim$0.5T). 
Fourier analysis of $\rho_{xx}(1/B)$, shown in Fig.~\ref{fig:Rxx}(c),  resolves only a single frequency $f$ peak at $0.25$T. The width of the peak, $\Delta f\sim0.09$T, is the level of uncertainty corresponding to a density difference $\Delta p$ less than $2\times10^{9}$ $cm^{-2}$. Therefore, the heavy hole (HH) band is approximately degenerate. The corresponding subband density $p=(g_s e/h)f=6\times10^{9}$ $cm^{-2}$ with ($g_s=2$), leading to a total $p$ of $1.2\times10^{10}$ $cm^{-2}$. 

\begin{figure} [b]
\includegraphics[totalheight=2.6in]{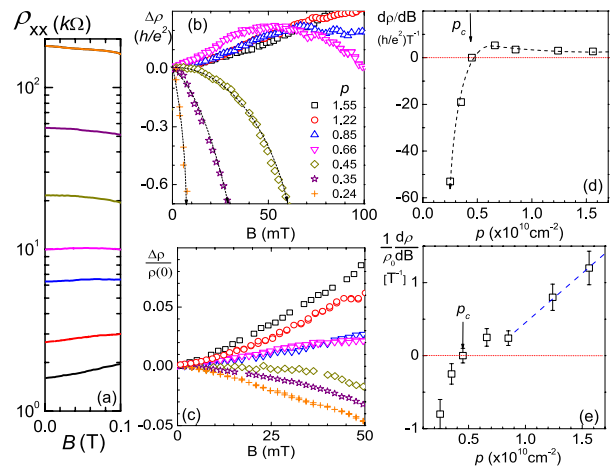}

\caption{\label{fig:weakB} (a) Low field magnetoresistance $\rho_{xx}(B)$ for various $p=$ (from top down) $0.24$, 0.35, 0.45, 0.66, 0.85, 1.22, and $1.55\times10^{10}$ cm$^{-2}$; (b) $\Delta\rho=\rho(B)-\rho(0)$ vs. $B$ for the same fixed $p$ shown in (a); (c) Results in (b) normalized by $\rho(0)$. (d) $d\rho/dB$ vs. $B$. (e) Results in (d) normalized by $\rho(0)$.}
\end{figure}

The correction $\Delta\rho$ to the MR for the degenerate HH band differs from the scenario of amply split subbands that possess not only unequal charge carrier masses (due to the warped dispersion caused by SOC), but also the non-proportionally different densities~\cite{SOI0,winkler}. Significantly split HH subbands exhibit different mobility $\mu_i$ (band index $i=1,2$) and the interband scattering gives rise to a correction $\Delta\rho(B)/\rho(0)\propto(\mu_1-\mu_2)^2$~\cite{interbandE}. However, for the (nearly) degenerate case, the two-band classical Drude term vanishes, so does the inter-band scattering since $\mu_1=\mu_2$. In addition, even the single-band Drude term is out, if isotropic scattering assumed, because the Lorentz force is cancelled by the Hall field~\cite{interbandE}. Therefore, $\Delta\rho(B)/\rho(B=0)$ here is mainly due to the quantum corrections, especially by SOC, as long as a Fermi Liquid or, more generally, $T<T_F$ is valid. As shown below, large $\Delta\rho$ in the $r_s\gg1$ limit grows progressively with decreasing $p$ (or increasing interaction) in a non-monotonic fashion as the system undergoes MIT.

Fig.~\ref{fig:weakB}(a) shows the MR for a series of $p$ from 0.2 to $1.5\times10^{10}$ $cm^{-2}$, approximately 1-5\% of the carrier densities used in Refs.~\cite{SOI1,SOI2}. $\rho$ reaches $h/e^2$ (or 25.8 k$\Omega$) around $p_c\sim 4\times10^{9}$ cm$^{-2}$ where the sign of MR switches. For $p>p_c$, the positive MR measures of the strength of SOC which produces corrections $\Delta\rho(B)=\rho(B)-\rho(0)$ (in the unit of $h/e^2$) dependent on $p$ [Fig.~\ref{fig:weakB}(b)]: larger $\Delta\rho(B)$ for lower $p$, in the opposite trend of decreasing SOC with lower $k_{\parallel}$~\cite{Rahimi}. The derivative $d\rho/dB$ shown in Fig.~\ref{fig:weakB}(d), taken within 30 mT, captures this increasing WAL effect with decreasing $p$ peaked around $6\times10^{9}$ $cm^{-2}$. Negative MR develops rapidly below $p_c$ and becomes stronger with decreasing $p$. Fig.~\ref{fig:weakB}(c) shows the same $\Delta\rho$ result in (b) normalized by $\rho(0)$. Taking the derivatives $d\rho/dB$ normalized by $\rho(0)$, a local maximum is again captured as shown in Fig.~\ref{fig:weakB}(e).

As $k_{\parallel}$ is significantly reduced, the band splitting $\Delta$ may exhibit a crossover from both the linear (BIA) and the cubic (SIA) dependence for larger $k_{\parallel}$ to the linear-$k_{\parallel}$ dependence for smaller $k_{\parallel}$~\cite{winkler}. Unlike the depletion cases, decreasing the gate voltage $V_g$ reduces the electric field according to $E=(V_g-V_{fb})/d=-\nabla V$ (where $V_{fb}$ is the flab-band voltage, $d$ is the barrier thickness, and $V$ is the approximately triangular confinement potential with a width $w$). The electric field is expressed as
\begin{equation}
E=\frac{2E_g ep}{\epsilon}   
\end{equation}
where $E_g$ is the triangular well height, $e$ the electron charge, and $\epsilon=13$ is the dielectric constant. For a 50\% decrease in $p$ from 1.55 to $0.85\times10^{10}$ cm$^{-2}$, $E$ is reduced by approximately $\sim$ 15\%. This agrees well with the gate voltage change from $V_g=-1.62$V to $V_g=-1.37$V corresponding to $p=$1.55 and $0.85\times10^{10}$ $cm^{-2}$ respectively. The well width $w=(2\epsilon E_g/ep)^{1/2}$ is widened by 1.4 times, resulting in a 30\%-decrease in the z-direction (growth direction) wave-vector $k_z=\pi/w$ of the Airy-like wavefunctions. The Dresselhaus coupling parameter $\beta\propto k_z^2$ is halved. If we consider the $k_{\parallel}^3$ splitting, it has a $p^{1.5}$ dependence since $p\sim k_{\parallel}^2/2\pi$ (2D). Since $\beta$ is also decreasing, the nearly linear decrease of $[1/\rho(0)]d\rho/dB$ (dotted line in Fig.~\ref{fig:weakB}(e)) with decreasing $p$ is reasonable. 

For this $p$ range, the splitting $\Delta$ of the HH band due to the $k_{\parallel}^3$ Rashba term should be considered. It has been shown for the accumulation-type heterostructures that the confinement electrical field $E$ is $\propto p$~\cite{stern}. Using the Airy wave-functions confined to a triangular well, the Rashba coefficient $\alpha\propto p^{-4/3}$~\cite{winkler}, which is enhanced 2.5 times when $p$ is reduced from 1.55 to $0.85\times10^{10}$ $cm^{-2}$. Because the observed overall correction is linearly decreasing for this $p$ range, it suggests a lesser Rashba effect than the Dresselhaus for higher $p$.  

Notice that for $p=$ 1.55, 1.22, and $0.85\times10^{10}$ $cm^{-2}$, the system is already strongly correlated with $r_s= 20\sim28$. It turns out the usual fitting of MR corrections based on the non-interacting model~\cite{HLN} already becomes problematic. Our fitting attempts produces 7 , 1.5, and 0.6 $ps$ respectively for $\tau_{so}$, however, with large errors ($R^2=0.94\sim0.89$). Note that $k_F l$, which is between 1 and 10, is much smaller than the cases in ref.~\cite{SOI2} and $\Delta\rho/\rho(0)$, amounting to 15\% at $B=50 m$T, is beyond perturbation. Clearly, a model appropriately incorporating interaction is needed. The theory on $\Delta\sigma(B)$ due to interaction by Altshuler et al~\cite{Altshuler} might be more relevant, even though it does not produce non-monotonic corrections for the large $r_s$ situations discussed below.

Remarkably, the decreasing trend in $[1/\rho(0)]d\rho/dB$ with lowering of $p$ is replaced by a striking 40\% increase (or 300\% in $d\rho/dB$) around $6\times10^{9}$ cm$^{-2}$. It signifies a substantially rising SOC coupling coefficient, particularly, the Rashba coefficient $\alpha$. In principle, as the $k_{\parallel}^3$ dependence fades at low $p$, the linear-$k_{\parallel}$ (or $p^{1/2}$) dependence eventually gives in to the rising $\alpha\propto p^{-4/3}$, resulting in the upturn of $[1/\rho(0)]d\rho/dB$. This is qualitatively in agreement with the data. There are other relevant effects that should also be recognized. Lowering $p$ to extremely dilute limits shrinks $k_{\parallel}$ to the HH-LH anti-crossing point~\cite{mixing} where the consequent effect on $\alpha$ is not fully known. Meanwhile, growing effective interaction with reducing $p$ raises $r_s$ beyond 30 and it gives rise to effects via exchange interaction~\cite{Utah} through an enhancement factor $\lambda_{SO}$. Theory predicts an approximately linear relation $\lambda_{SO}\sim r_s$ for moderate $r_s$ values less than 20~\cite{Utah}. Another interaction-driven effect derives from the enhancement of the effective mass $m^*$ which has been known to exist in the close vicinity of the $p_c$ of the MIT~\cite{AKS-review}. Due to the non-parabolic dispersion relation at low $p$ and the HH-LH mixing, $m^*$ is a complicated quantity. Actual comparison would require a measurement of $m^*$ which is difficult to achieve with cyclotron resonance (due to the small energy) or the SdH (due to the large Coulomb energy). Nevertheless, the non-monotonic rising is consistent with an enhanced SOC which influences the MR opposite to the trend of localization as $p$ crosses below $p_c$. Therefore, SOC supports MIT~\cite{pudalovSOC}.

$\Delta\rho(B)\rightarrow 0$ at $p=p_c$ corresponds to the diminishing WAL eventually overcome by the localization. A Wigner crystal (WC) regime is arrived at $p<p_c$ or $r_s>40$. The role of SOC depends on the actual carrier state which is likely a WC liquid because of the following: Due to the tiny energy scales, $E_F\sim 150-400$ mK, a WC is fragile to the influences of disorder, thermal~\cite{mermin-wagner} and quantum fluctuations (on a scale of $\sim$480 nm$/\sqrt{T}$). Consequently, the melting temperature $T_m$ is usually reduced well below the classical estimate: $\sim E_{ee}/130\sim 120 mK$. A recent study demonstrates a dynamical pinned WC marked by enormous pinning strength and extremely sharp dc-VI threshold~\cite{exp}. The $T$-dependence of pinning suggest a $T_m\sim 30$ mK which is lower than the temperature for the MR measurement. Meanwhile, since those data also support a second-order phase transition, an intermediate phase~\cite{spivak-kivelson} is relevant for the MR results for $T>T_m$. The quantum scenario still holds because $T$ is less than $T_F=E_F/k_B$, plasma frequency $\Omega=\sqrt{r_s}E_F$, and $E_{ee}$. Thus, the hydrodynamic flow model for the semiquantum case~\cite{andreev78,andreev11} does not apply. 

The negative MR for $p<p_c$ is intuitively surprising since an external $B$-field should stabilize a WC. On the other hand, the $B$-field raises the Zeeman energy $g\mu_B B$ which increases the overall carrier energy, favoring delocalization. Though, the Zeeman effect is small for higher $p$ (or weakly interacting) cases, it becomes prominent as the $g$-factor is greatly enhanced with increasing interaction (or lower $p$)~\cite{g-enhanced}. For a reducing $p$ (below $p_c$), both the absolute Coulomb energy $E_{ee}=e^2\sqrt{\pi p}/\epsilon$ and the $E_F=(\pi\hbar^2/m^*)p$ decrease. Moreover, for lower enough $k_{\parallel}$, 2D DOS diverges (towards the van Hove singularity), resulting in lower kinetic energy states. Thus, interaction effect, including the $g$-factor, is further enhanced.  As the Zeeman energy grows with increasing $B$, a larger negative MR for lower $p$ is a possibility. Relevant transport theory is needed.   

In summary, interaction-driven effects are realized in high purity ultra-dilute systems possessing unique transport behaviors distinct from the activated transport in Anderson insulators (or percolation) governed by disorder. SOC-driven band splitting is small in the dilute limit. Yet, there exists a substantial non-monotonic rise in the WAL in the metallic side, consistent with a delocalization effect. The switching of the signs of MR is in excellent correspondence to the zero-field MIT. The rising SOC coupling parameters are due to the large interaction including the scrambled exchange energy. The negative MR in the insulating side seems to be consistent with a rising carrier energy through the Zeeman term facilitated by the reducing absolute Coulomb energy. 

This work is supported through NSF-1105183 and NSF-1410302. The work at Princeton was partially funded by the Gordon and Betty Moore Foundation through Grant GBMF2719, and by the National Science Foundation MRSEC-DMR-0819860 at the Princeton Center for Complex Materials.

\bibliography{biblio.bbl}
\end{document}